\documentclass[twocolumn,showpacs]{revtex4-1}  
\usepackage{graphicx}

\usepackage{amsthm}
\usepackage{color}
\usepackage{amsmath}   
\usepackage{accents}
\usepackage{float}

\usepackage{subfigure} 
\usepackage{amssymb}    
\usepackage{graphicx}   
\usepackage{dsfont}
\usepackage{array}
\usepackage{epstopdf}
\usepackage{setspace}
\usepackage{tabularx}

\begin{document}
\title{From Metadynamics to Dynamics}
\author{Pratyush Tiwary}
\email{ptiwary@ethz.ch}
\thanks{Corresponding author}
\author{Michele Parrinello}
\affiliation{Department of Chemistry and Applied Biosciences, ETH Zurich, and Facolt\`a di Informatica, Istituto di Scienze Computazionali, Universit\`a della Svizzera Italiana Via G. Buffi 13, 6900 Lugano Switzerland}
\date{\today}

\newtheorem{thm}{Theorem}[section]
\newtheorem{prop}[thm]{Proposition}
\newtheorem{claim}[thm]{Claim}
\newtheorem{lem}[thm]{Lemma}
\newtheorem{cor}[thm]{Corollary}

\begin{abstract}

Metadynamics is a commonly used and successful enhanced sampling method. By the introduction of a history dependent bias which depends on a restricted number of collective variables(CVs) it can explore complex free energy surfaces characterized by several metastable states separated by large free energy barriers. Here we extend its scope by introducing a simple yet powerful method for calculating the rates of transition between different metastable states. The method does not rely on a previous knowledge  of the transition states or reaction coordinates, as long as CVs are known that can distinguish between the various stable minima in free energy space. We demonstrate that our method recovers the correct escape rates out of these stable states and also preserves the correct sequence of state-to-state transitions, with minimal extra computational effort needed over ordinary metadynamics. We apply the formalism to three different problems and in each case find excellent agreement with the results of long unbiased molecular dynamics runs.

\end{abstract}

\pacs{02.70.Ns, 05.70.Ln, 87.15.H-, 5.20-y}

\maketitle

Molecular dynamics (MD) simulation is a powerful and much used tool in many scientific fields. In spite of its many successes, MD is limited in scope by its inability to describe long time-scale dynamical processes. This can be a severe limitation since much interesting dynamics takes  place as the system moves from one free energy basin to another through infrequent rare events which can occur after waiting times  often well exceeding the millisecond time scale. On the other end, in fully atomistic simulations the integration time step needs to be of the order of femtoseconds to correctly integrate the equations of motion. This makes impractical in many cases to wait for the relevant rare events to take place spontaneously and  in spite of  the remarkable progress in purpose built computers \cite{deshaw}, the time scale problem still remains a serious issue.  While not much can be done about the integration timestep, there has been progress in developing enhanced sampling methods that can overcome these bottlenecks following different strategies \cite{meta_pnas,wanglandau,wtm,chandler_tse,chandler_pnas,voter_prl,flooding1,vangunsteren,gervasio_prl,milestoning,forwardflux,laio_cluster,abc}. While many of these methods have focused only on reconstructing the static properties, some have also tried to calculate dynamic properties \cite{chandler_tse,chandler_pnas,voter_prl,flooding1,vangunsteren,gervasio_prl,milestoning,forwardflux,laio_cluster,abc}. However, for various reasons the application of these methods has not been as widespread as one would hope for and there is a clear need for new and possibly simpler methods.

Here we shall take metadynamics \cite{meta_pnas} which is a successful enhanced sampling method used to calculate static properties, and show how it can be used to calculate dynamic properties in a simple way. In metadynamics one identifies a few collective variables (CVs), and then by depositing a history dependent biasing potential as a function of these CVs typically in the form of Gaussians \cite{vangunsteren}, the system is assisted in escaping free energy minima and visiting new regions in configuration space that would be practically inaccessible in unbiased MD. The efficiency of metadynamics in doing what it was primarily designed to do, namely recover free energy surfaces (FES) for complex systems, is by now well established \cite{meta_review}. So far though it has not been possible to estimate dynamic properties from these simulations, with the notable exception of Ref. \cite{laio_cluster} where a complex post-processing procedure relying on a number of assumptions has been suggested. 

More precisely, our aim is to obtain the correct sequence of state-to-state transitions, and to estimate the  time the system spends on average in each metastable state. In this letter we present and validate a powerful yet easy to use formalism that achieves these objectives while still maintaining full atomistic resolution. There is no extra computational effort needed as compared to ordinary metadynamics and in contrast with other methods \cite{laio_cluster,gervasio_prl} the post-processing is minimal. We are inspired by previous dynamical extensions of accelerated sampling methods \cite{voter_prl,flooding1} based on the addition of a static bias. However we show that we are able to avoid some serious limitations of these approaches. Most notably we do not need to know the location and nature of transition pathways or any of the reaction coordinates beforehand. We provide three different examples of increasing complexity that establish the validity of our approach.

An integral part of any metadynamics run is the choice of a small set of CVs  or descriptors \{$s_i({\bf R })$\} which are nonlinear functions of the atomic coordinates \textbf{R}. For simplicity in the following we denote the CVs as $s$. The CVs are able to distinguish between reactants and products and help sample different basins, but they are not required to form a basis for the ensemble of reaction pathways \cite{chandler_tse}.

{We suppose for argument's sake, that there exists a reaction coordinate $\lambda({\bf R)}$ such that for $\lambda\leq \lambda^*$ we are in the starting basin and for $\lambda>\lambda^*$ in a second basin, and the hypersurface  $\lambda({\bf R})=\lambda^*$ defines the dividing surface which also contains the transition state (TS) or equivalently the dynamical bottleneck for moving between the two basins\cite{hanggi,chandler_tse}. We assume that the time taken to cross this bottleneck is much less than the time spent in the individual basins, and that local equilibrium exists at all times.} We can then write the mean transition time $\tau$ over the barrier into the other state as:
\begin{equation}
\label{eq:tau0}
\tau = {1\over \omega \kappa} {Z_0 \over Z_0^*} =  {1\over \omega \kappa} {\int_{\lambda \leq \lambda^*} d\textbf{R} e^{-\beta U(\textbf{R})}\over \int_{\lambda=\lambda^*} d\textbf{R} e^{-\beta U(\textbf{R})}  } 
\end{equation}
{Here $\omega$ is a normalization constant detailed in Ref. \cite{doll,truhlar}. $\kappa$ is a transmission coefficient accounting for TS  recrossing events \cite{flooding1,chandler_tse, hanggi} whose value does not concerns us as we show soon for systems of interest in this letter where the transition through the bottleneck is fast.  For $\kappa=1$, Eq. (\ref{eq:tau0}) is equivalent to the result of transition state theory \cite{hanggi}. But no such assumption is needed here, neither do we need to calculate $\kappa$ itself which is another advantage over TST.} $Z_0$,  $Z_0^*$ are partition functions for the system confined to the first basin and to the hypersurface $\lambda=\lambda^*$ respectively with averages performed over the Boltzmann ensemble, $U({\bf R})$ is the interaction potential, and $\beta = {1\over k_B T}$ is the inverse of temperature multiplied by the Boltzmann constant $k_B$.

Let us now assume that we can perform a metadynamics run in which by accumulating bias against visited states we gradually enhance the probability of visiting $\lambda=\lambda^*$, but do not deposit bias over regions near the TS. {The bias is applied as a function of some CVs $s$ which are required to distinguish between the deep minima of the two basins. This is a much weaker requirement than on the order parameter $\lambda$ which should be able to identify the dynamical bottleneck as well.} As we show through our examples later, it is easier to find such CVs rather than the corresponding order parameter. The mean transition time $\tau_M(t)$ for the metadynamics run changes as the simulation progresses and is given by
\begin{equation}
\label{eq:taum}
\tau_M(t) = {1\over \omega \kappa_M} {Z_M(t) \over Z_M^*(t)}
\end{equation}
where $\kappa_M$, $Z_M$ and $Z_M^{*}$ are analogues of $\kappa$, $Z_0$ and $Z^*_0$ in Eq. (\ref{eq:tau0}), but are sampled using the time-dependent probability density of metadynamics \cite{wtm}.

If there is no bias deposited in the TS region around $\lambda=\lambda^*$, the dynamics of the system near it will be unaffected, implying $\kappa_M \approx \kappa$ and $Z_M^* \approx Z_0^*$.  Thus generalizing to metadynamics the results of  Ref. \cite{voter_jcp,flooding1} we write the acceleration factor $\alpha={{\tau} \over {\tau_M}}$ as:
\begin{equation}
\label{eq:alpha1}
\alpha(t) \approx {Z_0 \over Z_M} = \langle e^{\beta(V(s(\textbf{R}),t) )} \rangle_M
\end{equation}
where the angular brackets denote an average over a  metadynamics run  confined to $\lambda \leq \lambda^*$, and $V(s,t$) is the metadynamics time-dependent bias. In the above argument the crucial assumption is that in Eqs. (\ref{eq:tau0}-\ref{eq:taum}), only the denominators depend on the behavior in the TS region. Also, a precise knowledge of $\lambda^*$ is not necessary since the values of $Z_0$ and $Z_M$ are dominated by configurations deep inside the basin. Thus we expect this approach to work even in cases where there is an ensemble of transition states defined via committor analysis \cite{chandler_tse}. Ultimately the validity of Eq. (\ref{eq:alpha1}) stands on the dynamics being Markovian in nature \cite{hanggi}.

To  make practical use of Eq. (\ref{eq:alpha1}) and recover true time from metadynamics, we need to avoid depositing bias in the TS region and, in the lack of a precise knowledge of this region, have a way of recognizing whether it has been crossed. The first condition is  simply met by increasing the time lag between two successive Gaussian depositions.  Since in a rare event regime the time the system takes to cross the TS region is rather short \cite{chandler_tse}, it is most unlikely that the crossing of the barrier and the Gaussian deposition occur at the same time and we can rule out this circumstance. {Whether the Gaussian deposition is infrequent enough can be ascertained by performing a few simulations with increasingly slower deposition frequency until the transition times converge within desired accuracy.} Of course if we were to continue the run for a very long time, eventually we would deposit Gaussians in the TS region and metadynamics would reach its diffusive converged limit in which the FES is fully reconstructed. This is not our objective here and we are able to obtain converged rates much before this limit. 

To complete the algorithm, we need to recognize when the system has moved from one basin to another even if we do not know the corresponding TS precisely. For this we follow the evolution of the acceleration factor $\alpha(t)= {1\over t}\int_0^t dt' e^{\beta V(s,t') }$ estimated from the running temporal average over the metadynamics time $t$. The transition from one basin to the other is encoded in the time derivative of $\alpha(t)$:
\begin{eqnarray}
\label{eq:speedup}
{d \alpha \over d t} =   {1\over t} \left[ e^{\beta V(s,t) } -  {1\over t}\int_0^t dt' e^{\beta V(s,t') }  \right]
\end{eqnarray}
which  exhibits a clear kink whenever the system  crosses a barrier and enters a new state, since the first term in the bracket changes abruptly while the second one, which is a running average, changes much more slowly. 
\begin{figure}
\centering
\subfigure[ ]{
\includegraphics[scale=0.1]{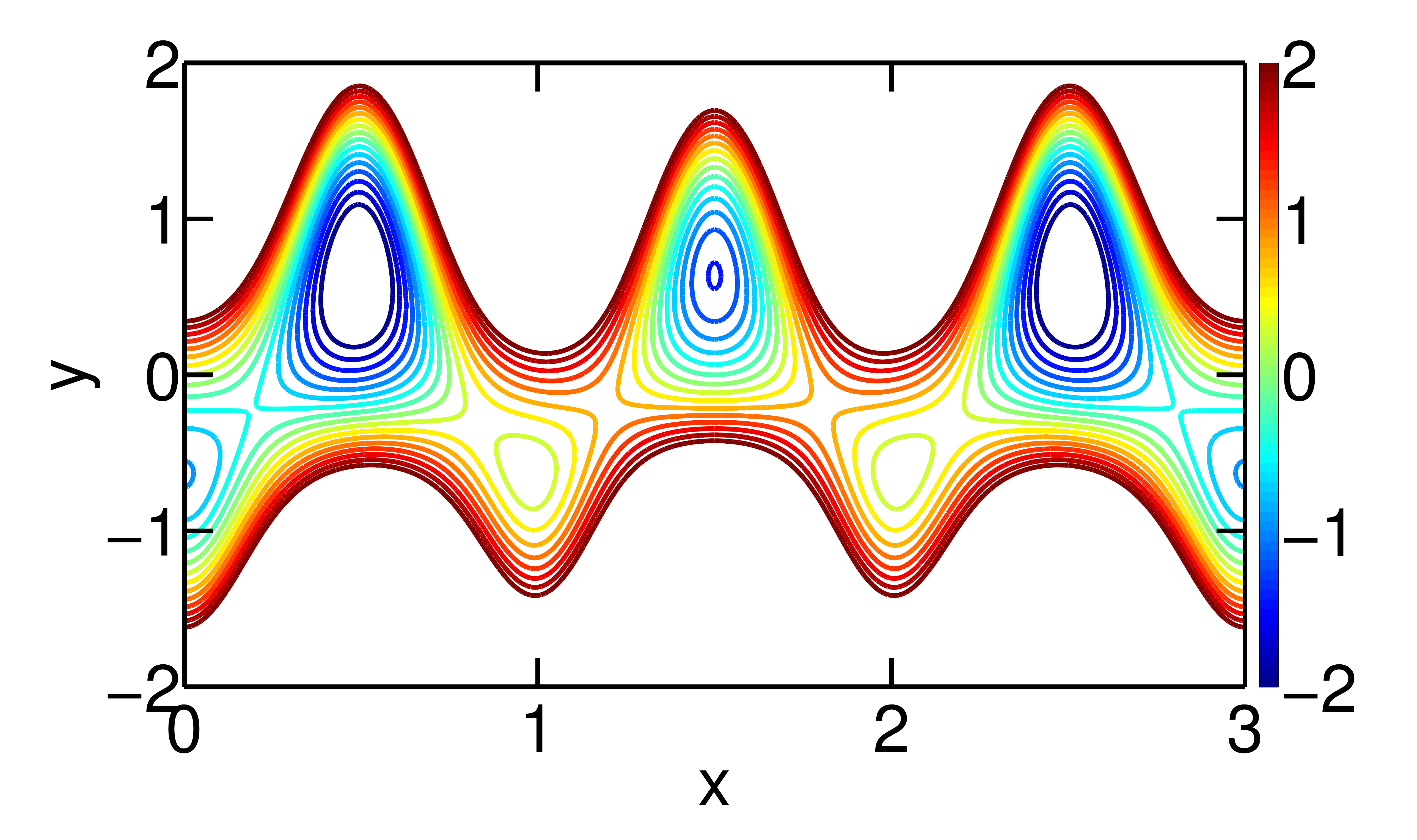}} 
\subfigure[ ]{
\includegraphics[scale=0.1]{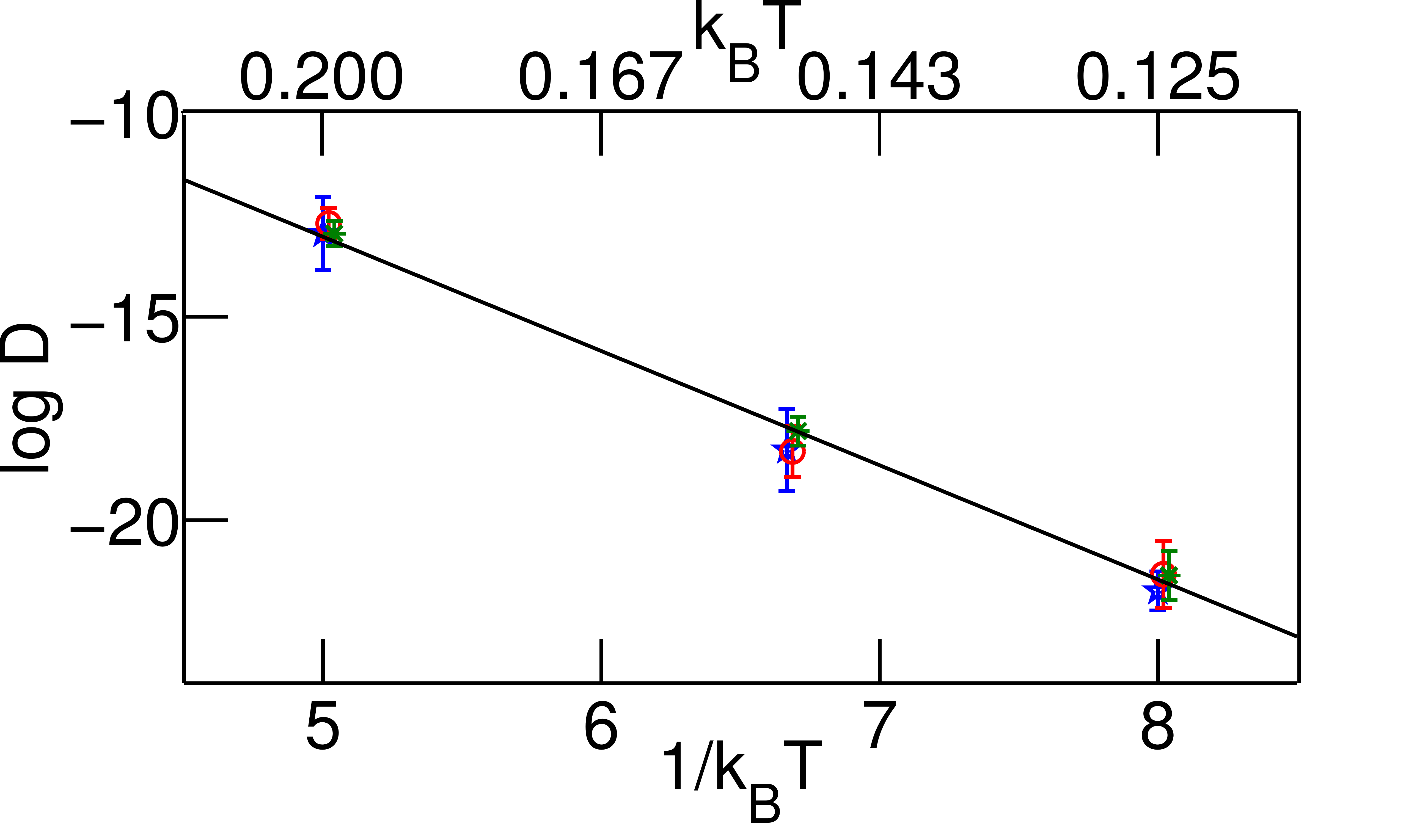}} 
\caption{(a) Model potential energy surface from Ref. \cite{voter_jcp} showing various stable states. The highest contour is at $V$=2 energy units and contours are separated by 0.25. (b) Logarithm of the one-dimensional diffusion constant versus inverse temperature (bottom) and temperature (top). The stars (blue), circles (red) and asterisks (green) correspond to calculations for three values of the well-tempered metadynamics effective temperature \cite{wtm,wte}: 0.75, 0.625 and 0.5 respectively, while the solid line is for kinetic Monte Carlo calculations reported in Ref. \cite{voter_jcp}. 95$\%$ confidence intervals are provided.}
\label{fig:voter}
\end{figure}
As we shall illustrate below with our examples (see Fig. \ref{fig:vacuum}(c) and supplemental information (SI)), this discontinuous change is easy to identify and gives us a clear one-dimensional marker for when the TS is crossed, irrespective of the number of CVs used. Clearly we do not know precisely when the system has crossed the watershed between the two minima, but as discussed earlier, this induces only a very small uncertainty since the time lag between depositions is a few picoseconds as compared to much longer transition times. We can also monitor if bias has been added to the TS region by overlaying the instants of bias deposition on an acceleration versus metadynamics time plot. One can then simply discard such a run. However we have not yet encountered such a case.  

We now proceed with a few illustrative applications of our approach. The calculations have been performed using standard simulation tools \cite{gromacs,plumed,amber,bussi_thermo} and the computational details can be found in SI. In SI  we also  provide  graphical evidence that, following our recipe of infrequent bias deposition, no bias was deposited in the TS regions. The first example is a 2-dimensional potential shown in Fig. \ref{fig:voter}(a).  This potential has multiple stable states connected through two pathways with different barriers and jump lengths. The long-time mean squared displacement and thus the diffusion constant depend on accurately sampling  both pathways. While it is easy to sample both at higher temperatures, at low temperatures the pathway with the higher barrier is rarely taken. By using metadynamics with potential energy as CV (the so-called well-tempered ensemble \cite{wte}) we can sample both pathways at all temperatures, and obtain accurate diffusion constants. Fig. \ref{fig:voter}(b) compares our results with the  calculations of Ref. \cite{voter_jcp} which come in part from direct MD simulation, and for the lower temperatures from kinetic Monte Carlo. It can be seen from Fig. \ref{fig:voter}(a) that potential energy is clearly nowhere near a good reaction coordinate for transitions in this system, yet the method works since this CV can distinguish between the various metastable states and well-tempered metadynamics favours transitions from state to state by enhancing the energy fluctuations \cite{wte,wtm}.

\begin{figure}
\subfigure[ ]{
\includegraphics[scale=0.085]{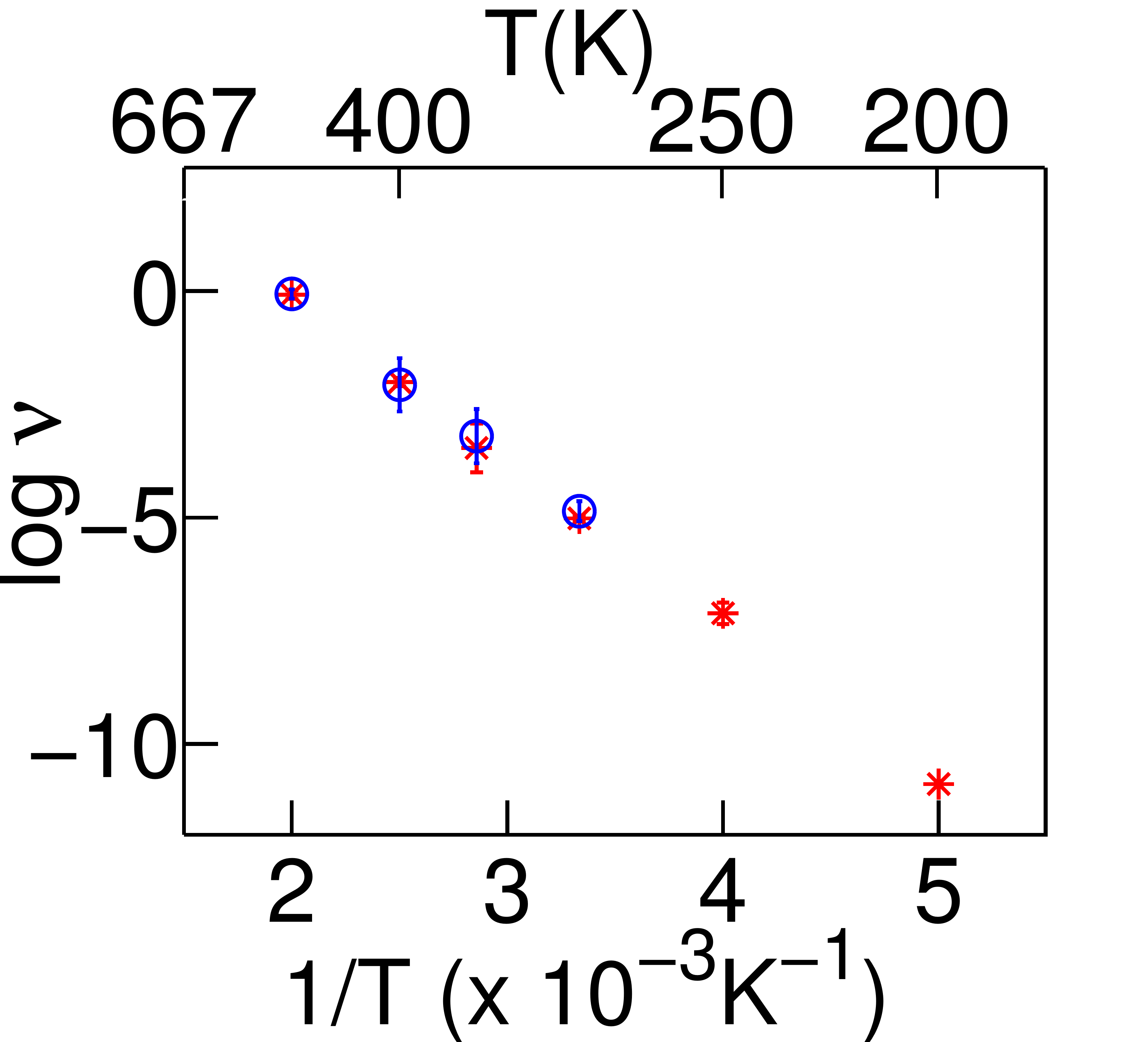}}
\subfigure[ ]{
\includegraphics[scale=.08]{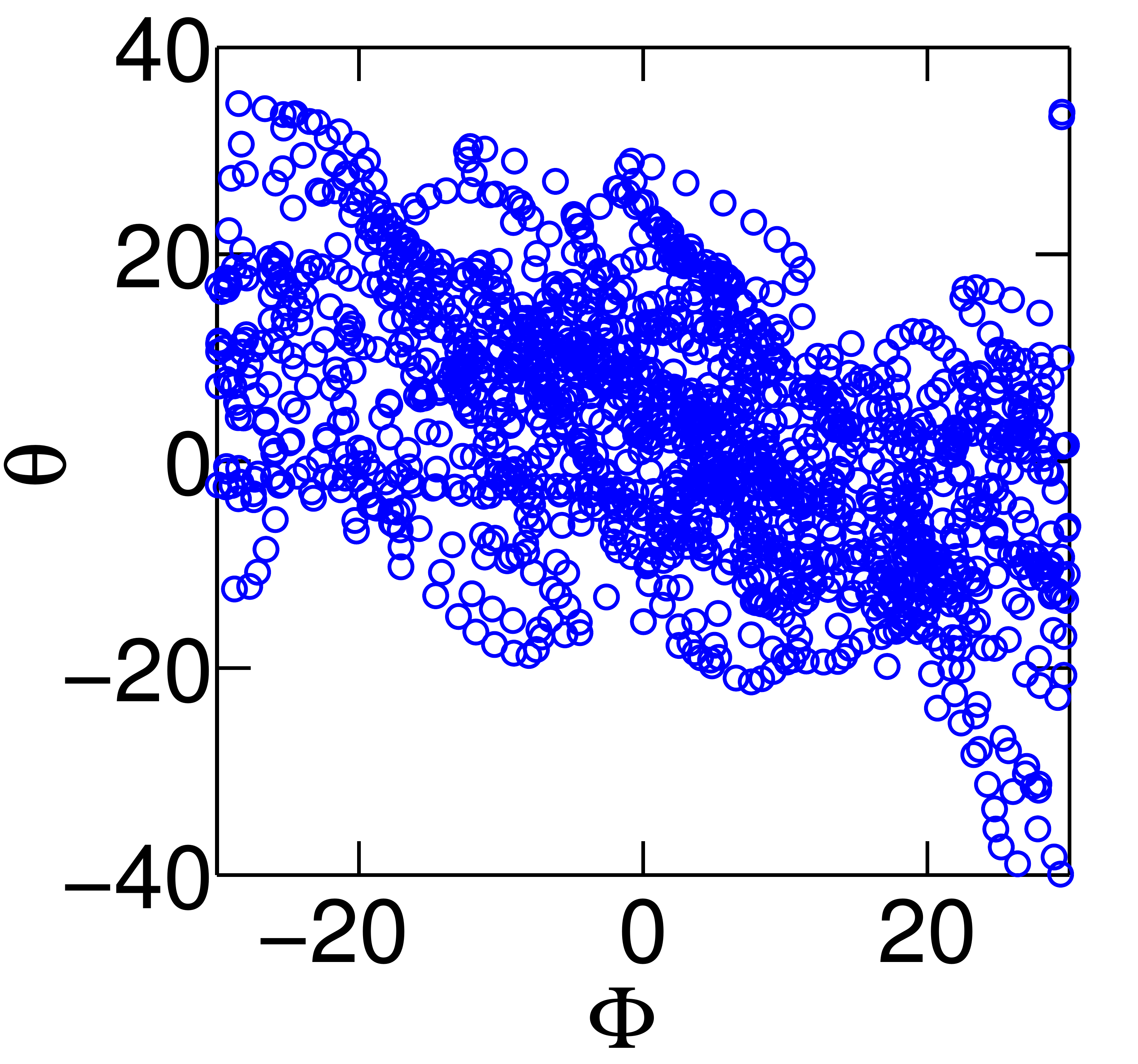}} 
\subfigure[ ]{
\includegraphics[scale=.10]{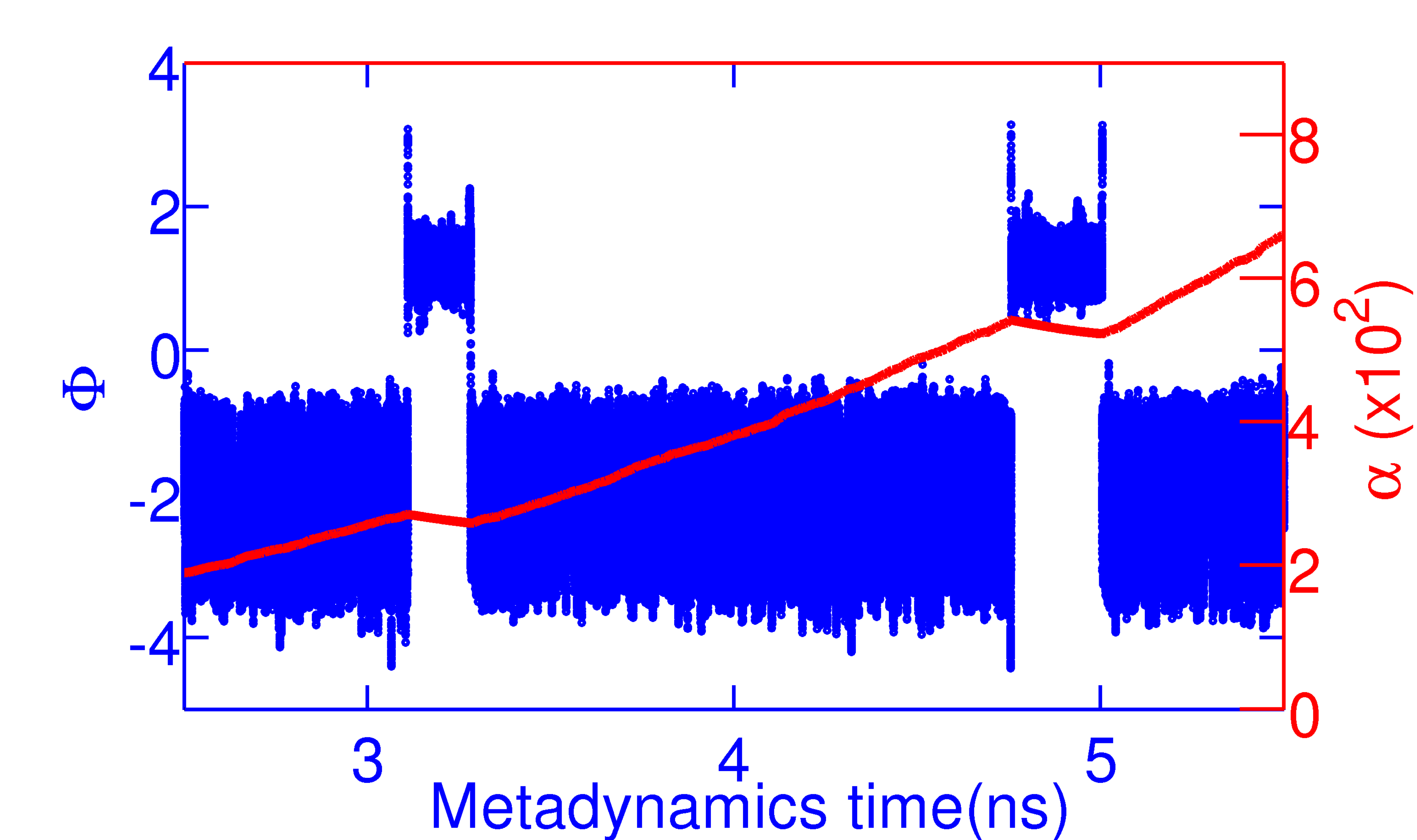}} 
\caption{(a) Logarithm of one over the average escape time (log $\nu$) from $C_{7eq}$ to $C_{7ax}$  versus inverse temperature (bottom) and temperature (top). Stars (red) are from metadynamics, circles (blue) are from long unbiased MD. Below 300 K, we did not observe transitions in unbiased MD within constraints of computer power. 95$\%$ confidence intervals are also provided. (b) The $\theta-\Phi$ anti-correlation in trajectories as they cross over from $C_{7eq}$ to $C_{7ax}$, obtained from metadynamics without using $\theta$ as a CV. (c) Acceleration (Eq. (\ref{eq:alpha1})) versus metadynamics time at T = 300 K, overlaid on the system's trajectory in the $\Phi$-dimension showing distinct kinks each time a TS is crossed.}
\label{fig:vacuum}
\end{figure}
\begin{table}
\caption{Transition rates($\nu$) for $\alpha \leftrightarrow \beta $ isomerization reaction for artificially stiffened alanine dipeptide molecule in water at 300 K, as obtained through unbiased MD and through metadynamics. 95$\%$ confidence intervals are also provided.}
\begin{ruledtabular}
\begin{tabular}{lll}
\textbf{Method}&$\nu_{\alpha \rightarrow \beta} (\mu sec ^{-1})$&$\nu_{\beta \rightarrow \alpha} (\mu sec ^{-1})$\\
unbiased MD&67 $\pm$ 27& 13 $\pm$ 5  \\
metadynamics&95 $\pm$ 18 & 16 $\pm$ 2  \\
\end{tabular}
\end{ruledtabular}
\end{table}
The second example is the $C_{7eq} \rightarrow C_{7ax}$ conformational change of alanine dipeptide  in vacuum. These two stable states can be distinguished by the values of the backbone dihedral angles ($\Phi,\Psi $) and are separated by a barrier of $\approx$ 8 kcal/mol (see SI for FES and dihedral angles definitions). Due to the high barrier, this has been a standard test system for many rare events methods \cite{meta_pnas,wte,wtm,bolhuis_jcp}. Here energy is not able to distinguish between the minima (the difference in energies of the minima is only $\approx$ 2 kcal/mol \cite{wtm}), thus we performed well-tempered metadynamics simulations using $\Phi$ and $\Psi$ as CVs. In Fig. \ref{fig:vacuum}(a), the so-obtained frequencies for $C_{7eq} \rightarrow C_{7ax}$ isomerization across various temperatures are compared to values obtained from long unbiased MD runs, and the agreement is near perfect. It is well known that a third dihedral angle $\theta$ is also part of the reaction coordinate \cite{chandler_pnas}. We show in Fig. \ref{fig:vacuum}(b) that even though we did not include $\theta$ as a CV in our simulations, we find a clear $\theta-\Phi$ anti-correlation in the trajectories that cross over from $C_{7eq}$  to  $C_{7ax}$. This anti-correlation is an essential feature of the TS ensemble, as found through detailed transition path sampling calculations \cite{chandler_pnas}. This once again brings forth a key feature of our approach. As long as we know CVs that can demarcate stable states and push the system out of basins, we do not need to identify other CVs that might be involved in the TS ensemble. Fig. \ref{fig:vacuum}(c) shows the acceleration as a function of metadynamics time, superimposed on the  $\Phi$ trajectory. A sharp change in slope can be seen each time a TS is crossed. The overall speed up of our calculation with respect to brute force MD is more than three orders of magnitude.

As a  third example, we consider again alanine dipeptide, but this time in water at T = 300 K. The FES for this system has been studied in Ref. \cite{bolhuis_jcp}, and with a barrier of 2 kcal/mol \cite{meta_pnas}, it is not exactly a rare event system since the $\alpha \leftrightarrow \beta $ isomerization frequency is on the order of tens per nanosecond, and thus an accelerated sampling approach is not needed. Thus we artificially stiffened the torsion terms in the force field \cite{amber} in order to make it a rare event system but still one in which we could  ascertain the effect of solvent's presence. Table I compares the values obtained through metadynamics (using $\Phi$ and $\Psi$ as CVs) to those from unbiased MD. The method is again rather  accurate, while providing an acceleration of 3 to 4 orders of magnitude. Kinks similar to those in Figs. \ref{fig:vacuum}(c) were found each time a transition occurred (see Fig. 3 in SI).  This result is very encouraging since it says that even in the presence of a fluctuating environment our method is expected to work.

As stressed earlier, expressions similar to Eq. (\ref{eq:alpha1}) have already been used in the literature in connection with fixed bias simulations 
\cite{voter_prl,flooding1,bondboost,sisyphus1,sisyphus2}. In these methods, it is required \textit{a priori} to have a sense of the nature of transitions and locations of various TS, and then construct a time-independent biasing potential that leaves these TS unperturbed. This is feasible if the FES and the relevant reaction coordinates are known. But in a complex system  it is  difficult to obtain all this information and in particular to identify the exact location and distribution of the TS. This has restricted the applicability of these methods, especially to systems where many degrees of freedom are simultaneously at play, and the notion of the TS itself has to be replaced by a whole ensemble of likely transition pathways (the TS ensemble) \cite{chandler_tse, chandler_pnas}. 

We circumvent this difficulty with our simple procedure. No assumption is made other than the quasi-stationarity of  metadynamics, a low residence time in TS regions and the use of a set of CVs that can help sample correctly the stable states of the system. The growing metadynamics literature provides examples of  rather generic CVs that have been successfully applied to a large variety of systems. The same CVs can now be used to extract rates. Since no bias has been added to the transition states, the system evolves with a state-to-state sequence that is preserved from the unbiased dynamics \cite{voter_jcp}.  If more precise information on the TS ensemble is needed, a committor analysis \cite{chandler_pnas} can be performed starting from the reactive paths harvested  in our simulations. {The method is designed for systems with rare-but-fast transitions where the time for crossing dynamical bottlenecks is small. As such it might not work efficiently for mesa-like barriers where the system spends a long time in the barrier region itself\cite{milestoning,forwardflux}. Nevertheless, our preliminary investigations on a variety of complicated systems with aptly chosen CVs are very encouraging regarding the range of applicability of our method.} We expect this new method to be extremely useful for calculating kinetic pathways and rates for a variety of complex systems, complementing the already established ability of metadynamics to calculate free energy surfaces.

The authors thank David Chandler for theoretical insight, and Ali Hassanali and Matteo Salvalaglio for useful and stimulating discussions.This work was funded by the European Union (Grant ERC-2009-AdG-247075). Supercomputer time was provided by CSCS on the s309 project and by ETH Zurich.
	\bibliography{draft0}
\end{document}